\documentstyle[11pt,newpasp,twoside,epsf]{article}
\def\eg{e.g., }
\def\etal{et~al.\ }
\def\edcomment#1{\iffalse\marginpar{\raggedright\sl#1\/}\else\relax\fi}
\reversemarginpar

\begin{document}
\title{Results from a diffuse intracluster light survey}
 \author{John J. Feldmeier, J. Christopher Mihos, Heather L. 
Morrison, Paul Harding, \& McBride, C.}
\affil{Case Western Reserve University, 10900 Euclid Ave., 
Cleveland OH, 44106, USA}

\begin{abstract}
We give an update of our ongoing survey for intracluster light (ICL),
in a sample of distant Abell clusters.  We find that the amount of
intracluster starlight is comparable to that seen in nearby clusters,
and that tidal debris appears to be common.  
\end{abstract}

\section{Introduction}

The concept of intracluster starlight (ICL), or stars between the galaxies
in galaxy clusters is not a new one: it was first proposed over
50 years ago (Zwicky 1951).  However, progress in studying ICL
has been slow due to its low surface brightness, which is less 
than 1\% of the brightness of the night sky.
This is unfortunate, because ICL is a powerful probe of the evolution 
of galaxies in clusters (Dressler 1984), and of cluster evolution overall.
It is also an important part of the ejection of matter out
of galaxies discussed at this conference.  

We have undertaken a deep imaging
survey of galaxy clusters, intended to quantify the properties of ICL
as a function of environment, and overall galaxy cluster properties.
From our deep imaging, with careful attention to systematic errors 
(\eg Morrison, Boroson, \& Harding 1994), we are able 
to measure the ICL to faint surface brightnesses many magnitudes 
below that of the night sky ($\mu_{\mbox{v,ICL}}$ $\approx$ 26--28).  
In tandem with the observations, we are constructing numerical 
simulations of galaxy clusters in a cosmological context, similar
to those of Dubinski (1998).  Here we summarize some recent results:
previous results can be found in Feldmeier \etal (2002).

\section{Intracluster Tidal Debris}
Models of intracluster star production predict that intracluster
light may not always be in a smooth, elliptically symmetric,
component but instead can be in tidal tails and arcs 
(Moore \etal 1996; Napolitano \etal 2003).  In our deep imaging
survey, we have found intracluster tidal debris in almost every
cluster we have surveyed (see Figure~1 for one of the most
striking examples).  However, at this time, we have detected more
plume-like structures than the long arcs originally predicted by
the models.  Since only half of the survey has been completely
reduced, the significance of this result is unclear. 

\section{The amount of intracluster light }

Determining the amount of ICL from our imaging observations 
is non-trivial because it is difficult to separate the  
outer edges of galaxies from the underlying background ICL.
To make a model-independent estimate of ICL fraction in 
the clusters observed, we used an isophotal cutoff.  
The exact value of the cutoff is problematic, so we
have adopted a range of values to set reasonable limits.
We find that for the clusters surveyed thus far, the amount
of ICL is 10--20\%, including a correction for the edges
of galaxies still unmasked.
This is similar to the amount of intracluster light
found in the nearby Virgo and Fornax clusters 
(see Arnaboldi, this volume).

\begin{figure}
\plotfiddle{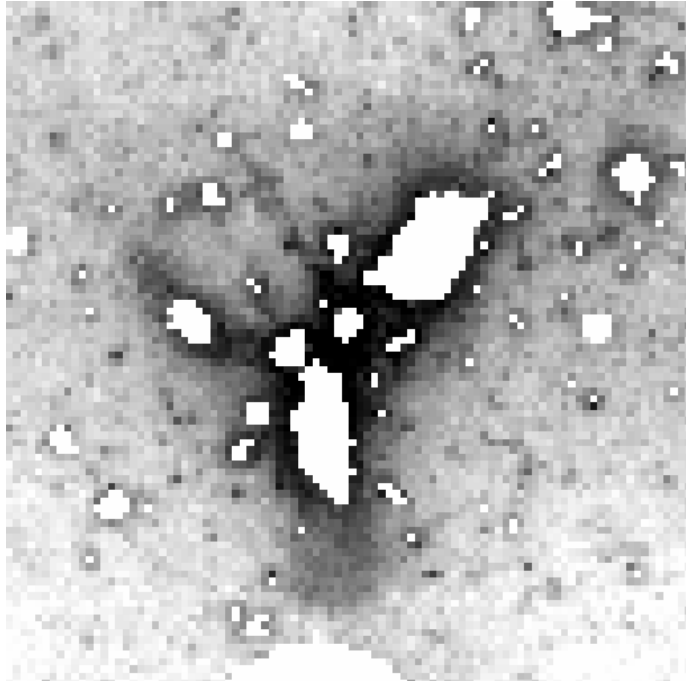}{3.3in}{0.}{50.}{50.}{-150}{-70}
\caption{An image of the central region of Abell~1914, with the
galaxies partially masked out, and the image binned into 11 x 11
pixel bins to show detail.  There is a large fan-shaped plume at
the bottom of the image, as well as multiple other low surface
brightness features.}

\end{figure}

\end{document}